\documentstyle[twoside,natbib,epsf]{article}

\input{ibvs2.sty}

\def\cite#1{\citealt{#1}}

\def\ibvs{Inf. Bull. Var. Stars}

\def\apj{ApJ}
\def\apjl{ApJ}

\def\apss{Ap\&SS}
\def\aap{A\&A}

\def\mnras{MNRAS}
\def\nat{Nature}
\def\pasp{PASP}

\def\VSOLJBul{VSOLJ Var. Star Bull.}
\def\PublisherCambridge{Cambridge: Cambridge University Press}

\begin{document}

\IBVShead{5243}{8 March 2002}

\IBVStitle{Unusual Outbursting State of a Z Cam-Type Star HL CMa}

\IBVSauth{Kato,~Taichi$^1$}
\vskip 5mm

\IBVSinst{Dept. of Astronomy, Kyoto University, Kyoto 606-8502, Japan,
          e-mail: tkato@kusastro.kyoto-u.ac.jp}

\IBVSobj{HL CMa}
\IBVStyp{UGZ}
\IBVSkey{dwarf nova}

\begintext

   HL CMa is well-renowned dwarf nova which was discovered as an Einstein
X-ray source (\cite{fuh80hlcma}; \cite{chl81hlcma}; \cite{bai81hlcma};
\cite{mei81hlcma}).  Although the possible Z Cam-type nature had long
been suggested (cf. \cite{man94hlcmasscyg}), the exact classification
of the object required more than a decade until the detection of an
unmistakable long standstill in 1999 (\cite{wat00hlcma}).  Several
authors reported relatively unusual spectroscopic features in HL CMa
(e.g. \cite{war83hlcma}).  Although later observations could not confirm
the result (\cite{cro86dqherpolari}), there was even a claim of the
possible presence of circular polarization (\cite{chl81hlcma}).
The object was thus intensively observed, particularly in the ultraviolet
(\cite{bon82hlcmav1223sgr}; \cite{mau87hlcmaIUEapss,mau87hlcmaIUE}),
which revealed the presence of significant outflow.
\citet{sti99hlcma} further studied the system, and obtained an orbital
period of 0.2146 or 0.2212 d.  In spite of relatively rich observations
in the past, no spectroscopic observation during standstills has been
reported, presumably because of the rarity of standstills.

   The typical outburst cycle length of HL CMa is 15 d (cf. \cite{chl81hlcma};
see also Figure 1).  During standstills, this outburst pattern disappears
(Figure 2) as in other Z Cam stars (cf. \cite{war95book}).

\IBVSfig{9cm}{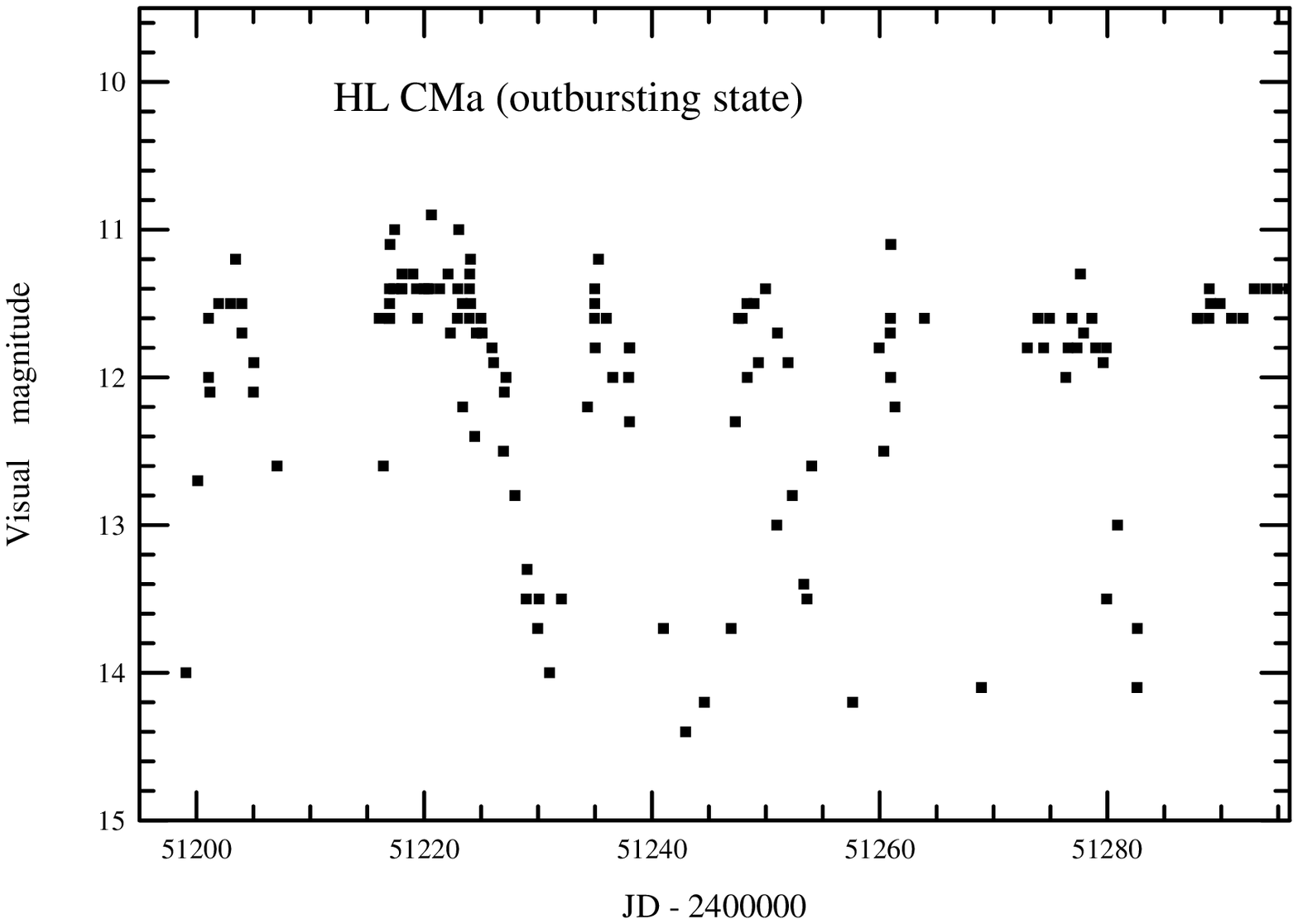}{
    HL CMa in normally outbursting state.  The data are from visual
observations reported to VSNET (http://www.kusastro.kyoto-u.ac.jp/vsnet/).
The errors of visual observations are usually less than 0.5 mag, which
do not affect the discussion.  Outbursts recur every $\sim$15 d.
}

\IBVSfig{9cm}{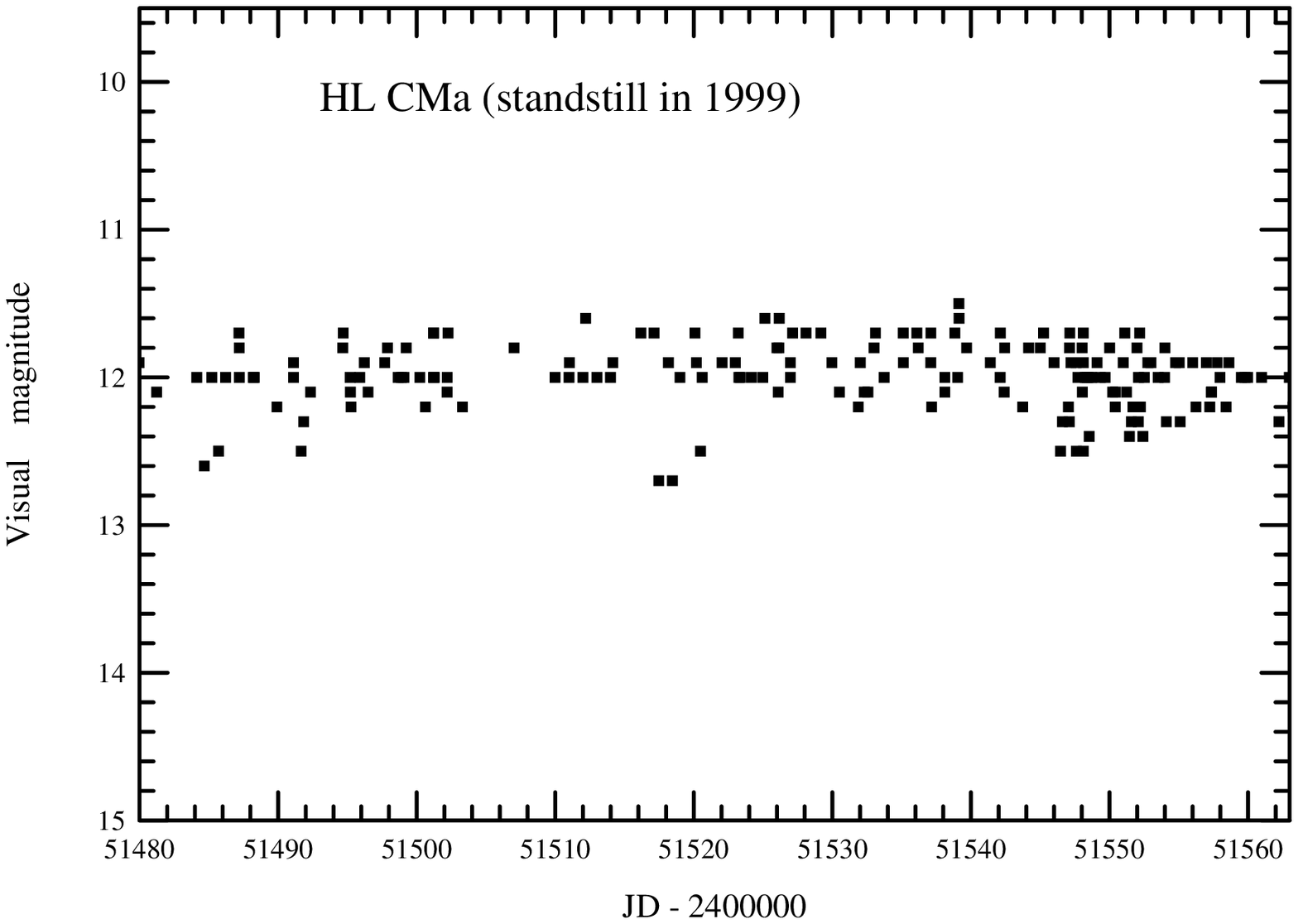}{
    Representative light curve of the standstill of HL CMa in 1999
(data from VSNET).  The star stopped outbursting.
}

   In 2001--2002, we noticed the presence of ``the third" outbursting
state (Figure 3).  During this period, the star showed weak ($\sim$1 mag)
outbursts with a longer ($\sim$30 d) outburst cycle length.
The outburst amplitude was intermediate between that of normally outbursting
state (Figure 1) and nearly zero during standstills (Figure 2).
Although the decrease of the outburst amplitude could be a result of
an increased mass-transfer rate, the lengthening of the cycle length
is quite unexpected, because an increase of mass-transfer rate generally
leads to a shortening of the cycle length (e.g. \cite{can88outburst},
in which HL CMa was listed as an object already close to the instability
border).

\IBVSfig{9cm}{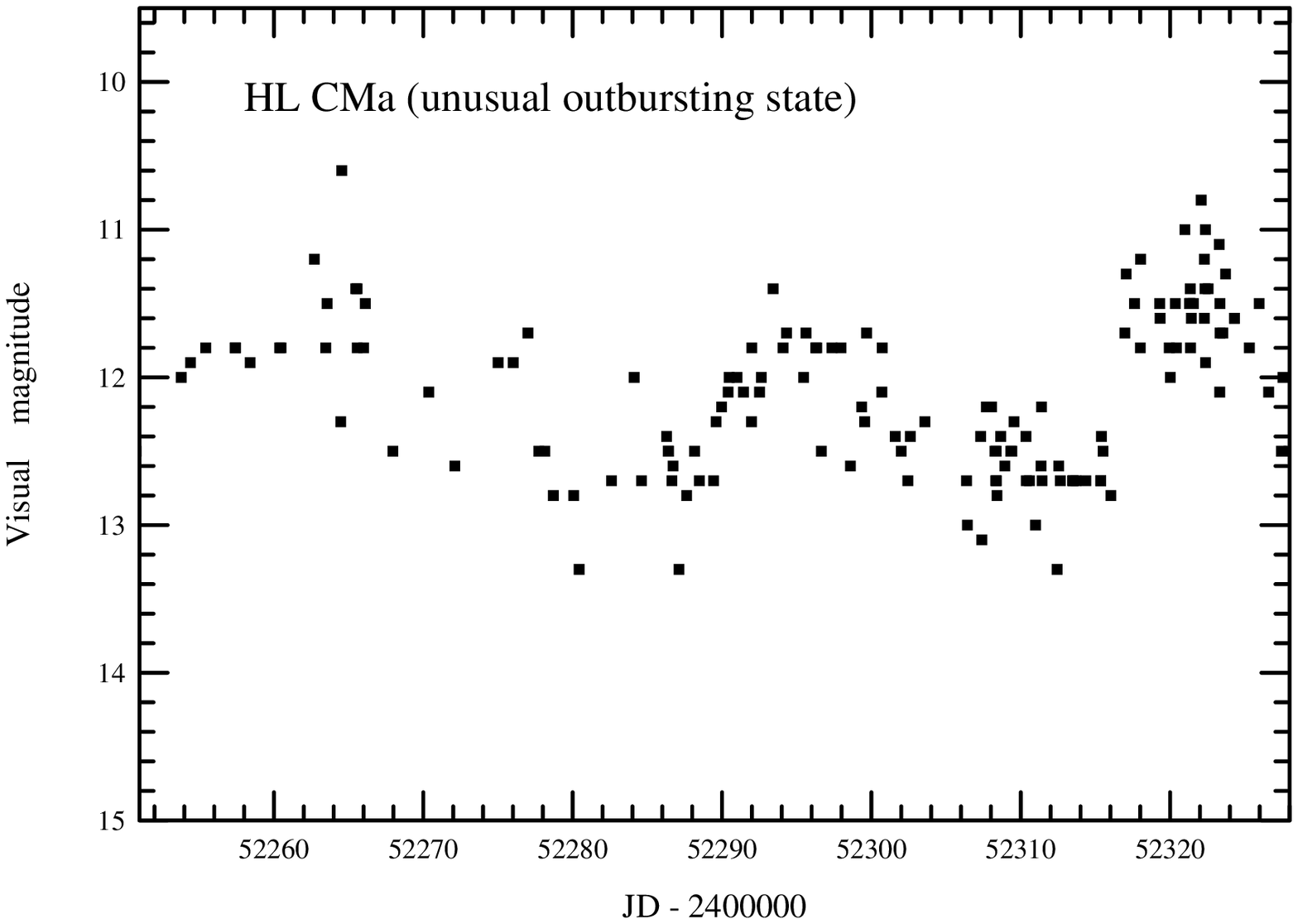}{
   Light curve of ``the third state'' of HL CMa in 2001--2002.
The star showed weak ($\sim$1 mag)
outbursts with a longer ($\sim$30 d) outburst cycle length.
}

   Such behavior may be compared to an unusual slow fading of a standstill
in another Z Cam-type star AT Cnc (\cite{kat01atcnc}).
\citet{kat01atcnc} proposed that this behavior may be a combined result
of heating on the accretion disk at an accretion rate slightly below
the stability, analogous to fadings of VY Scl-type stars
(\cite{lea99vyscl}).  The presence of
strong P Cyg feature in the ultraviolet (\cite{bon82hlcmav1223sgr};
\cite{mau87hlcmaIUEapss}) and the unusual presence of high-excitation
optical lines (\cite{war83hlcma}; \cite{chl81hlcma}) could be interpreted
as an emerging signature of strong irradiation field.  Since there have
been only few occasions of unusually outbursting states in the
decades-long history,
X-ray and spectroscopic observations to detect further signatures of
high-energy photons and irradiation are highly encouraged during the
present unusual state.

\vskip 3mm

We are grateful to many VSNET observers who have reported vital observations.
This work is partly supported by a grant-in aid (13640239) from the
Japanese Ministry of Education, Culture, Sports, Science and Technology.

\end{document}